\documentclass{aa} % for a referee version
\usepackage{graphicx}
\usepackage{epsfig} 
\usepackage{graphicx}
\usepackage{amssymb}
\usepackage{natbib}
\usepackage{amsmath}
\def\her{Her\,X-1}
\def\0115{4U\,0115+63}
\def\cen{Cen\,X-3}
\def\1626{4U\,1626-67}
\def\lmc{LMC\,X-4}
\def\exo{EXO\,2030+375}
\def\a0535{A\,0535+262}
\def\4u1907{4U\,1907+09}
\begin{document}

\title{Can disk-magnetosphere interaction models and beat 
frequency models for quasi-periodic oscillation in accreting X-ray pulsars be reconciled?} 

\author{E. Bozzo\inst{1,2}\fnmsep\thanks{\email{bozzo@oa-roma.inaf.it}} 
\and L. Stella\inst{1}
\and M. Vietri\inst{3} 
\and P. Ghosh\inst{4}} 

\offprints{E. Bozzo}
\titlerunning{Threaded disk models in X-ray pulsars}
\authorrunning{E. Bozzo et al.}

\institute{INAF - Osservatorio Astronomico di Roma, Via Frascati 33,
00044 Rome, Italy 
\and Dipartimento di Fisica - Universit\`a di Roma "Tor Vergata", via
della Ricerca Scientifica 1, 00133 Rome, Italy 
\and Scuola Normale Superiore, Pisa, Italy 
\and Department of Astronomy and Astrophysics, Tata Institute of Fundamental Research, Mumbai 400 005, India}

\abstract{}{In  this paper we review some aspects of the theory of magnetic threaded disks.} 
{We discuss in particular the equations that determine the position of the inner disk boundary   
by using different prescriptions for the neutron star-accretion disk interaction.  
We apply the results to several accretion powered X-ray pulsars that showed both 
quasi-periodic oscillations in their X-ray flux and spin-up/spin-down torque reversals.  
Under the hypothesis that the beat-frequency model is applicable to the quasi-periodic 
oscillations, we show that these sources provide an excellent opportunity to test models 
of the disk-magnetosphere interaction.}  
{A comparison is carried out between the magnetospheric radius obtained with all the prescriptions used 
in threaded disk models; this shows that none of those prescriptions is able to reproduce the 
combination of quasi-periodic oscillations and torque behaviour observed for different X-ray luminosity levels in the 
X-ray pulsars in the present sample.} 
{This suggests that the problem of accretion disk 
threading by stellar magnetic field is still lacking a comprehensive solution.  
We discuss briefly an outline of possible future developments in this field.} 


\date{Received 07-24-2008 / Accepted: 25-10-2008}

\maketitle

\section{Introduction}
\label{sec:intro}
The problem of the interaction between the magnetic field of a 
neutron star (NS) and the surrounding accretion disk has been investigated 
by a number of studies, with the aid 
of magnetohydrodynamic simulations \citep{lamb,sharl,ghoshlamb,ghoshlamb0,
ghoshlamb1,ghoshlamb2,lovelace,romanova,romanova1,romanova3,usty06}. 
Despite several aspects of disk-magnetosphere interaction are still poorly understood,  
the idea that the NS magnetic field must penetrate to some extend into the accretion 
disk (due to instabilities leading to finite conductivity of the plasma) is now widely accepted. 
This ``magnetically threaded disk model'', first developed in detail by 
\citet{ghoshlamb0}, \citet{ghoshlamb1,ghoshlamb2} 
and later revised by \citet{wang87,wang95}, predicts that,  
as a result of the NS magnetic field threading the disk, 
a magnetic torque is generated that exchanges angular momentum between the NS and the disk.  
The strength of this magnetic torque increases in the disk regions closest to the NS and exceeds the 
viscous stresses at the magnetospheric radius, $R_{\rm M}$, 
where the disk is terminated. 
The expression  
\begin{equation}
R_{\rm M}^{\rm sp}=\mu^{4/7}\dot{M}^{-2/7} (2GM)^{-1/7},  
\label{eq:rm}
\end{equation}
which holds strictly only for spherically symmetric accretion, 
can be considered only a rough approximation when the accretion flow is mediated by a disk   
\citep[in the equation above $M$ and $\mu$ are the neutron star mass and magnetic moment,  
and $\dot{M}$ is the mass accretion rate,][]{lamb}.  

In Sect. \ref{sec:review} we review theories of the threaded disk model.   
Particular attention is given to the different magnetic torque 
prescriptions of \citet{ghoshlamb1,ghoshlamb2} and 
\citet{wang87,wang95}, and the calculation of the magnetospheric radius in the 
two cases (hereafter GLM and WM, respectively).  
In Sect. \ref{sec:application} these calculations are applied to a sample of accretion powered 
X-ray sources that displayed both spin-up/spin-down torque reversals and quasi-periodic 
oscillations \citep[QPO,][]{klis95} in their X-ray flux. 
The X-ray luminosity at the onset of the spin-up/spin-down transition is used to fix poorly 
known parameters in the equations of the magnetospheric radius in both the GLM and WM, whereas 
the observed QPO frequencies are assumed to match the predictions of the beat frequency model 
(BFM, see Sect. \ref{sec:BFM}) in order to derive an independent estimate of $R_{\rm M}$.   

We carry out in Sect.~\ref{sec:discussion} a comparison between the magnetospheric radii  
obtained within the threaded disk models and the BFM, and show that neither the GLM nor the 
WM are able to reproduce observations for the whole sample of X-ray pulsars considered here. 
We argue that the problem of the threaded accretion disk might still 
lack a more general and comprehensive solution, and provide an outline of a  
revision of the GLM that will be presented in a 
subsequent paper (Bozzo et al., 2009, in preparation).

\section{A Review of the Magnetic Threaded Disk Model}
\label{sec:review}
The magnetic threaded disk model was developed by \citet{ghoshlamb1,ghoshlamb2} 
and partly revised by \citet{wang87,wang95}, under the assumption that the NS is rotating 
about its magnetic axis and that this axis is perpendicular to the plane of the disk 
(the so called ``aligned rotator'').   
The model is based on the idea that the stellar magnetic field must to some extend 
penetrate the accretion disk due to a variety of effects that prevent this field from being 
completely screened from the disk. 
Once this occurs, the differential motion between the disk, rotating at the Keplerian rate   
$\Omega_{\rm k}$, and the star, rotating with angular frequency $\Omega_{\rm s}$, 
generates a toroidal magnetic field, $B_{\phi}$, 
from the dipolar stellar field component
\begin{equation}
B_z(R)=-\eta\frac{\mu}{R^3}
\label{eq:bz} 
\end{equation}
($\eta$ is the screening coefficient, see below). 
The shear amplification of $B_{\phi}$ occurs on a time scale 
$\tau_{\rm d}$$\sim$$\vert$$\gamma$($\Omega_{\rm s}$-$\Omega_{\rm k}$)$\vert$$^{-1}$.  
Here $\Omega_{\rm k}$=$(GM/R^3)^{1/2}$, $\Omega_{\rm s}$=2$\pi$/$P_{\rm s}$, 
$P_{\rm s}$ is the spin period of the NS, $h$ is the disk height, 
$R$ is the radial distance from the NS, and 
$\gamma$$\simeq$1 parametrizes the steepness of the vertical transition between Keplerian 
motion in the disk and rigid corotation with the NS. 
The finite disk plasma conductivity leads to slippage of field lines through the plasma, and thus to 
reconnection of these lines above and below the symmetry plane of the disk. 
This opposes to the shear amplification of the toroidal magnetic field on a time scale  
$\tau_{\phi}$$\sim$$h$/($\xi$$v_{A\phi}$). Here the term $\xi$$v_{A\phi}$ defines the reconnection 
velocity in terms of the local Alfv\'en speed $v_{A\phi}$ and $\xi$$\sim$0.01-0.1, 
if the main dissipation effect is the annihilation of the poloidal field near the disk midplane, or  
$\xi$$\sim$1, if magnetic buoyancy is considered. \citet{ghoshlamb1} and 
\citet{wang87,wang95} proposed different prescriptions for $B_{\phi}$, 
and in the following we discuss their models separately. 

\subsection{The Ghosh \& Lamb model}
\label{sec:glreview} 
\citet{ghoshlamb1} estimated the toroidal magnetic field by equating the 
amplification and reconnection time-scales, {\it i.e.}  
\begin{equation}
\frac{B_{\phi}}{B_z}\simeq\mp \frac{\gamma
(\Omega_s-\Omega_k)h}{\xi v_{Az}}.  
\label{eq:alf1}
\end{equation} 
In their model, the coupling between the NS and the disk occurs in a broad transition zone 
located between the flow inside the NS magnetosphere and the unperturbed disk flow. 
The transition zone comprises two different regions: 
\begin{itemize}
\item The boundary layer, that extends from the inner disk boundary 
$R_{\rm M}^{gl}$ inward to a distance $\Delta_{\rm R}$$\ll$$R_{\rm M}^{gl}$. 
In the boundary layer the poloidal field is twisted at a 
level $B_{\rm \phi}$/$B_{\rm z}$$\sim$1, the angular velocity of the disk plasma 
significantly departs from the Keplerian value, and matter leaves the disk plane in the vertical direction  
and accretes onto the star. The currents generated on the disk surface by the magnetic 
field line twisting lead to an $\sim$80\% screening of the NS magnetic field ($\eta$$\sim$0.2 outside 
the boundary layer).    
\item The outer transition zone, where the disk flow is only slightly 
perturbed and the coupling between the disk and the star is provided by the residual 
NS magnetic field that survives the screening of the boundary layer. 
Despite this is only a small fraction of the NS magnetic field ($\sim$20\%), 
the broader extension of the outer transition zone makes this coupling appreciable. 
\end{itemize}

In this model the magnetospheric radius coincides with the outer radius of the boundary layer,  
and is defined as the radius at which viscous torques in the disk 
are balanced by magnetic torques produced by field line twisting within the boundary layer. 
\citet{ghoshlamb1} found 
\begin{equation}
R_{\rm M}^{\rm gl}\simeq0.52 R_{\rm M}^{\rm sp},   
\label{eq:rmgl}
\end{equation}  
where $R_{\rm M}^{\rm sp}$ is given by Eq.~\ref{eq:rm}. 

The total torque $N$ on the star depends on  
the torque $N_{\rm 0}$=$\dot{M}(GMR_{\rm M}^{\rm gl})^{1/2}$, produced by matter leaving the 
disk at R$_{\rm M}^{\rm gl}$ and accreting onto the NS, and the torque 
$N_{\rm mag}$=$-\int^{R_{\rm s}}_{R_{\rm M}^{\rm gl}}B_{\phi}B_{z}R^2 dR$ 
generated by twisted magnetic field lines threading the disk outside $R_{\rm M}^{\rm gl}$. 
Expressed in an adimensional form this torque is  
\begin{equation}
n=N/N_0=(N_0+N_{\rm magn})/N_0=1-\frac{\int^{R_{\rm s}}_{R_{\rm M}^{\rm gl}}B_{\phi}B_{z}R^2 dR}
{\dot{M}(GMR_{\rm M}^{\rm gl})^{1/2}}.  
\label{eq:gltorque} 
\end{equation}
Beyond $R_{\rm s}$, the outer radius of the transition zone, the NS magnetic field is completely 
screened by the disk, and no torque is produced.   
Note that the poloidal magnetic field $B_{\rm z}$ in Eq.~\ref{eq:gltorque} differs from 
the simple dipolar approximation of Eq.~\ref{eq:bz}. In fact, \citet{ghoshlamb1} calculated 
this component taking into account the effect 
of the screening currents flowing on the disk surface (see their Eq.~40). As a result, \citet{ghoshlamb2} 
found that $n$ is primarily a function of the so called ``fastness parameter''\footnote{The explicit functional form of 
$n(\omega_{\rm s})$ is given by Eq.~7 in \citet{ghoshlamb2}.}  
\begin{equation}
\omega_{\rm s}^{\rm gl}=\Omega_{\rm s}/\Omega_{\rm k}(R_{\rm M}^{\rm gl})=(R_{\rm M}^{\rm gl}/R_{\rm co})^{3/2},   
\label{eq:omegas}
\end{equation} 
where R$_{\rm co}$=1.5$\times$10$^8$ cm (M/M$_{\odot}$)$^{1/3}$  
$P_{\rm s}$$^{2/3}$ is the corotation radius. 
For a fixed mass accretion rate, 
the torque $n$ of Eq.~\ref{eq:gltorque} can be either positive (the NS spins up) or negative (the NS spins down), 
depending on the NS spin period.  
In particular, for slow rotators ($\omega_{\rm s}$$\ll$1) $n$$\simeq$1.4, and the star 
spins up, while increasing $\omega_{\rm s}$, $n(\omega_{\rm s})$ first 
decreases and then vanishes for the critical value $\omega_{\rm s}$=$\omega_{\rm c}^{\rm gl}$. 
For $\omega_{\rm s}$$>$$\omega_{\rm c}^{\rm gl}$, $n$ becomes negative and the star rotation  
is slowed down by the interaction with the accretion disk, whereas for 
$\omega_{\rm s}$$\gtrsim$0.95 no stationary solution exists and   
steady state accretion is not permitted. 

The critical fastness parameter $\omega_{\rm c}^{\rm gl}$ depends on the 
magnetic pitch angle at the inner disk radius, $\gamma_{0}$=B$_{\rm \phi}
(R_{\rm M}^{\rm gl})$/B$_{\rm z}(R_{\rm M}^{\rm gl})$; 
\citet{ghoshlamb2} suggested $\gamma_{0}$$\simeq$1, which corresponds to 
$\omega_{\rm c}^{\rm gl}$$\simeq$0.35 (see their Fig.~4). 
A critical fastness parameter much smaller than unity implies 
the torque on the NS is zero only when the magnetospheric radius is 
well inside the corotation radius and close to the compact star 
(see also Sect.~\ref{sec:application}).

\subsubsection{The magnetospheric radius in the Ghosh \& Lamb model}
\label{sec:magngl}

In the GLM, the magnetospheric radius is given by Eq.~\ref{eq:rmgl}. 
We define the variable $x_{\rm gl}$=R$_{\rm M}^{\rm gl}$/R$_{\rm co}$=$(\omega_{\rm s}^{\rm gl})^{2/3}$,  
and rewrite Eq.~\ref{eq:rmgl} in an adimensional form 
\begin{equation}
x_{\rm gl}=2.17 \mu_{30}^{4/7} \dot{M}_{16}^{-2/7} m^{-10/21} P_{\rm s}^{-2/3}. 
\label{eq:xgl}
\end{equation}

Here $m$ is the NS mass in units of 1M$_{\odot}$.  
Equation~\ref{eq:xgl} gives the ratio $x_{\rm gl}$ between the magnetospheric and  
corotation radii, for fixed values of $\mu_{30}$, $m$, $P_{\rm s}$, 
and $\dot{M}_{16}$. Using the definition of the critical fastness 
parameter and defining $x_{\rm c}^{\rm gl}$=($\omega_{\rm c}^{\rm gl}$)$^{2/3}$,  
Eq.~\ref{eq:xgl} translates into 
\begin{equation}
x_{\rm gl}=x_{\rm c}^{\rm gl} (\dot{M}_{16}/\dot{M}_{\rm tr16})^{-2/7},  
\label{eq:xglcri}
\end{equation}
such that the only free parameter is the 
mass accretion rate $\dot{M}_{\rm tr16}$ at which $x_{\rm gl}$=$x_{\rm c}$ 
({\it i.e.} the mass accretion rate at which the torque $n$ undergoes a sign reversal).    

In the GLM the magnetospheric radius can be easily estimated from Eq.~\ref{eq:xglcri}, 
as a function of the mass accretion rate $\dot{M}_{16}$, provided the only free parameter 
$\dot{M}_{\rm tr16}$ is somehow constrained by observations. 
This is discussed in Sect.~\ref{sec:application}.

\subsection{The Wang Model}
\label{sec:wmodel}

\citet{wang87} suggested that the toroidal field of Eq.~\ref{eq:alf1}  
is overestimated, as the magnetic torque diverges in the 
limit $R_{\rm s}$$\to$$\infty$. Instead of balancing the two time 
scales $\tau_{\rm \phi}$ and $\tau_{\rm d}$, he introduced a different 
prescription for the toroidal magnetic field based on 
Faraday's induction law. 
Assuming that the growth of the toroidal magnetic field is limited by 
reconnection in the disk (see also Sect. \ref{sec:glreview}), he found that 
\begin{equation}
\frac{B_{\phi}}{B_z}\simeq\pm\left\vert\frac{\gamma
(\Omega_s-\Omega_k)h}{\xi v_{Az}}\right\vert^{1/2},   
\label{eq:alf2} 
\end{equation}
and proved the amplification of the toroidal 
field to be smaller than previously thought (note that Eq. 
\ref{eq:alf2} is the square root of Eq. \ref{eq:alf1}).

In a later study, \citet{wang95} considered also that the growth of the toroidal field 
$B_{\phi}$ might be limited by other mechanisms than magnetic 
reconnection.  
In case the amplification of the toroidal field is 
damped by diffusive decay due to turbulent mixing within the disk,    
$\tau_{\phi}$=$(\alpha\Omega_k)^{-1}$ \citep[$\alpha$ is the viscosity 
parameter of][hereafter SS73]{ss73} and Eq. \ref{eq:alf2} is replaced by 
\begin{equation}
\frac{B_{\phi}}{B_z}\simeq\frac{\gamma(\Omega_s-\Omega_k)}{\alpha\Omega_k}.
\label{eq:alf3} 
\end{equation}

Another possibility is that, in the case of a force-free magnetosphere, the winding 
of the field lines threading the disk is limited by magnetic reconnection taking 
place within the magnetosphere itself. In this case 
\begin{equation}
\frac{B_{\phi}}{B_z}=
\left\{
\begin{array}{lr}
\gamma_{max}(\Omega_{\rm s}-\Omega_{\rm k})/\Omega_{\rm k}, 
&\Omega_{\rm k}\gtrsim\Omega_{\rm s};\\
\gamma_{max}(\Omega_{\rm s}-\Omega_{\rm k})/\Omega_{\rm s},  
&\Omega_{\rm k}\lesssim\Omega_{\rm s},  
\end{array}
\right.
\label{eq:alf4} 
\end{equation}
where $\gamma_{max}\lesssim1$.\\
At odds with the GLM, the model developed by \citet{wang87,wang95} 
does not involve the presence of a boundary layer: the effect 
of the screening currents is not taken into account self-consistently, and 
$B_{\rm z}$ is described by Eq.~\ref{eq:bz}, by assuming a constant screening $\eta$$\lesssim$1  
from the inner disk radius, $R_{\rm M}^{\rm w}$, up to the external boundary 
of the disk (taken to be at infinity).  
Accordingly, the total adimensional torque onto the NS is 
\begin{equation}
n=1-\frac{\int^{\infty}_{R_{\rm M}^{\rm w}}B_{\rm \phi} B_{\rm z}R^2 dR}
{\dot{M}(GMR_{\rm M}^{\rm w})^{1/2}}, 
\label{eq:wtorque} 
\end{equation}
where $B_{\rm \phi}$ is given by Eqs.~\ref{eq:alf2}, or \ref{eq:alf3} 
or \ref{eq:alf4}, and $R_{\rm M}^{\rm w}$ is derived from the equation 
\begin{equation}
\frac{B_{\rm \phi}(R_{\rm M}^{\rm w}) B_{\rm z}(R_{\rm M}^{\rm w})}
{\dot{M}(GMR_{\rm M}^{\rm w})^{1/2}}
=-\frac{1}{2(R_{\rm M}^{\rm w})^3}. 
\label{eq:rmw}
\end{equation}
Equation~\ref{eq:rmw} expresses the balance between the rates at which the stellar magnetic field 
and viscous stresses remove angular momentum from the disk. 
 
The torque given by Eq.~\ref{eq:wtorque} is positive for slow rotators 
($\omega_{\rm s}$$\ll$1) and negative in the opposite limit, in agreement 
with the results found in the GLM. 
However, the torque vanishes for critical values of the 
fastness parameter in the 0.88-0.95 range, {\it i.e.} well above the value $\omega_{\rm c}^{\rm gl}$=0.35 
predicted by \citet{ghoshlamb1}. In particular, $\omega_{\rm c}^{\rm w}$=0.949, 0.875, and 0.95  
for $B_{\phi}$ given by Eqs.~\ref{eq:alf2}, \ref{eq:alf3}, and \ref{eq:alf4}, 
respectively. A similar value is found in appendix \ref{app:A}, where we calculate 
the value of $\omega_{\rm c}$ for region ''C'' of a SS73 disk, 
as opposed to region ``B'' used by \citet{wang87}. 

At odds with the GLM, such large values of the critical fastness parameters in the WM 
imply that the magnetospheric radius must lie close to the corotation radius when 
$n$$\simeq$0. Therefore, NS spin-down can take place over a tiny range of mass 
accretion rate. This conclusion turns out to be nearly independent of the 
prescription used in the WM for the toroidal field. 
Taking these results into account, \citet{wang96} proposed that a 
constant magnetic pitch at the inner disk boundary, {\it i.e.} 
$\gamma_0=B_{\phi 0}B_{z0}$, might be assumed in Eq.~\ref{eq:rmw},  
and derived the simplified expression for the magnetospheric radius 
\begin{equation}
R_{\rm M}^{\rm w}=\xi_{\rm p} R_{\rm M}^{\rm sp}.  
\label{eq:rmwapprox}
\end{equation}
Here $\xi_{\rm p}$$\simeq$1.35$\gamma_0^{2/7}$$\eta^{4/7}$, 
$\eta$ is the screening factor of Eq.~\ref{eq:bz}, 
and $R_{\rm M}^{\rm sp}$ is given by Eq. \ref{eq:rm}. 

In Sect.~\ref{sec:magnet}, we solve Eq.~\ref{eq:rmw} 
numerically and compare the results 
with those obtained by using a constant pitch angle approximation.  
\begin{table*}
\scriptsize 
\centering
\caption{Properties of accretion powered X-ray pulsars discussed in the text}
\begin{tabular}{cccccccccc}
\hline
\hline
\noalign{\smallskip}
\tiny
Source & $P_{\rm s}$ & $\nu_{\rm s}$ & $\nu_{\rm QPO}$ & $L_{\rm QPO}$ & 
$L_{\rm tr}$ & $x_{\rm QPO}^{\rm BFM}$ & $x_{\rm QPO}^{\rm w}$ & $x_{\rm QPO}^{\rm gl}$ & Classification$^{a}$ \\
& & & & (band) & (band) & & & \\
\hline
\noalign{\smallskip}
& s & mHz & mHz & erg s$^{-1}$ & erg s$^{-1}$ &  & & &\\
& & & & (keV) & (keV) & & & &\\
\noalign{\smallskip}
\hline
\noalign{\smallskip}
\her $^{15,16,17}$ & 1.24 & 806.5 & 8 & $2.1\times10^{37}$ & $2.1\times10^{37}$ & 0.95-0.98 & 0.91-0.97 & 0.5 & F\\
& & & 43 & $(0.1-200)$ & $(0.1-200)$ & & & & \\
\noalign{\smallskip}

\0115\ $^{9,10,18}$ & 3.62 & 276.2 & 62 & $8\times10^{37}$ & $5\times10^{36}$ & 0.56 & 0.61-0.71 & 0.23 & F\\
& & & & $(0.1-200)$ & $(0.1-200)$ & & & &\\
\noalign{\smallskip}

\cen\ $^{13,14,19}$ & 4.8 & 208.3 & 35 & $ 1.0\times10^{38}$ & $3.3\times10^{37}$ & 0.89 & 0.82-0.92 & 0.36 & F\\
& & & & $(0.12-100)$ & $(0.12-100)$ & & & &\\
\noalign{\smallskip}

\lmc\ $^{11,12}$ & 13.5 & 74.1 & 0.65-1.35 & $10^{39}$ & $2\times10^{38}$ & 0.97-0.98 & 0.76-0.88 & 0.31 & F\\
& & & & $(2-25)$ & $(2-25)$ &  & & &\\
\noalign{\smallskip}

\1626\ $^{6,7,8}$ & 7.66 & 130.6 & 40 & 1.36$\times$10$^{36}$ & 1.7$\times$10$^{36}$ & 0.83 & 0.98-0.99 & 0.88 & F/S\\
& & & & $(0.7-60)$ & $(0.7-60)$ & & & &\\
\noalign{\smallskip}

\exo\ $^{1,2}$ & 42 & 23.8 & 213 & 2$\times$10$^{38}$ & $10^{38}$& 0.22 &  0.86-0.94 & 0.4 & S\\
& & & & $(1-20)$ & $(1-20)$ & & & &\\
\noalign{\smallskip}

               & 42 & 23.8 & 213 & 2$\times$10$^{38}$ & $2.4\times10^{36}$ & 0.22 &  0.43-0.56 & 0.14 & S\\
& & & & $(1-20)$ & $(1-20)$ & & & &\\
\noalign{\smallskip}

\a0535\ $^3$ & 103 & 9.7 & 72 & $4.3\times10^{37}$ & $4.83\times10^{36}$ & 0.24 & 0.69-0.82 & 0.26 & S\\
& & & & $(20-100)$ & $(20-100)$ & & & &\\
\noalign{\smallskip}

\4u1907\ $^{4,5}$ & 440 & 2.3 & 55 & $6.3\times10^{36}$ & $2\times10^{36}$ & 0.12 & 0.81-0.92 & 0.36 & S\\ 
& & & & $(2-60)$ & $(1-15)$ & & & &\\
\noalign{\smallskip}
\hline
\hline
\noalign{\smallskip}
\noalign{\smallskip}
\noalign{\smallskip}
\multicolumn{10}{l}{\textit{a:} F=Fast rotator, S=Slow rotator.} \\
\multicolumn{10}{l}{\textit{References:} (1) \citet{parmar}; (2) \citet{angelini}; (3) \citet{fin96}; 
(4) \citet{fritz}; (5) \citet{zand98}}\\
\multicolumn{10}{l}{(6) \citet{cha97}; (7) \citet{cha}; (8) \citet{shi}; (9) \citet{tam}; (10) \citet{camp01}}\\
\multicolumn{10}{l}{(11) \citet{woo}; (12) \citet{moon}; (13) \citet{howe}; (14) \citet{take}; (15) \citet{dalfiume98}}\\
\multicolumn{10}{l}{(16) \citet{par99}; (17) \citet{boroson00}; (18) \citet{soong89}; (19) \citet{burderi00}}\\
\end{tabular}
\label{tab:source}
\end{table*}

\subsubsection{The magnetospheric radius in the Wang model}
\label{sec:magnet}

Here we solve Eq.~\ref{eq:rmw} for the three different prescriptions 
of the toroidal magnetic field discussed in the previous section.  
We consider first the prescription of Eq.~\ref{eq:alf2}. 
In this case a model for the region of the accretion disk 
that is just outside the magnetosphere is required to evaluate 
the disk height $h$ and Alfv\'en velocity $v_{\rm Az}$.  
In accordance with \citet{wang87} we use the thin disk model of 
SS73 \citep[see e.g.,][]{vietri}.   
Using the well known relation $h$=$c_{\rm s}^2/\Omega_{\rm k}$ connecting the disk 
vertical height $h$ to the sound speed $c_{\rm s}$=($P$/$\rho$)$^{1/2}$  
(here $\rho$ the matter density and $p$ the thermal pressure inside the disk), 
Eq.~\ref{eq:alf1} translates into \citep{wang87} 
\begin{equation}
\frac{B_{\phi}}{B_z}\simeq\pm\left\vert\frac{\gamma(\Omega_s-\Omega_k)}
{\xi \Omega_k}\frac{(4\pi p)^{1/2}}{B_z}\right\vert^{1/2}.
\label{eq:alfpr}
\end{equation}
Introducing the expressions for the thermal pressure in the ``B'' and ``C''  
regions of the SS73 accretion disk, we obtain  
\begin{eqnarray}
x_{\rm w}^{-211/80}\sqrt{1-x_{\rm w}^{3/2}}& = & \nonumber 
 2.72\times10^{-3}\sqrt{\xi\gamma^{-1}\eta^{-3}}
\alpha^{9/40}\cdot\\ 
& &\cdot\mu_{30}^{-3/2}
\mu_p^{1/4}m^{7/6} P_{\rm s}^{211/120}\dot{M}_{16}^{4/5}, 
\label{eq:xalfb}
\end{eqnarray}
and
\begin{eqnarray}
x_{\rm w}^{-85/32}\sqrt{1-x_{\rm w}^{3/2}}&=& \nonumber 
 3.184\times10^{-3}\sqrt{\xi\gamma^{-1}\eta^{-3}}
\alpha^{9/40}\cdot\\ 
&& \cdot\mu_{30}^{-3/2}
\mu_p^{1/4}m^{7/6} P_{\rm s}^{85/48}\dot{M}_{16}^{63/80}, 
\label{eq:xalfc}
\end{eqnarray}
respectively. In the following we refer to these models as 
WM1 and WM2, respectively. Here $\dot{M}_{16}$ is the mass accretion rate 
in units of $10^{16}$ g s$^{-1}$, $\mu_{30}$ is the magnetic 
moment of the neutron star in units of $10^{30}$ G cm$^{-3}$,  
$\mu_{\rm p}$ is the mean molecular weight, and, in analogy to Sect.~\ref{sec:magngl}, 
we introduced the adimensional variable 
$x_{\rm w}$=R$_{\rm M}^{\rm w}$/R$_{\rm co}$. 
With similar calculations, we find  
\begin{eqnarray}
x_{\rm w}^{-7/2}-x_{\rm w}^{-2}& = & 
2.38\times10^{-3}\alpha\gamma^{-1}\eta^{-2}\mu_{30}^{-2}\cdot \nonumber \\ 
&& \cdot m^{5/3} P_{\rm s}^{7/3}\dot{M}_{16} 
\label{eq:xdiff} 
\end{eqnarray}
and
\begin{eqnarray}
x_{\rm w}^{-7/2}-x_{\rm w}^{-2}& = & 
2.38\times10^{-3}\gamma_{max}^{-1}\eta^{-2}\mu_{30}^{-2}\cdot \nonumber \\
&& \cdot m^{5/3} P_{\rm s}^{7/3}\dot{M}_{16}, 
\label{eq:xrec} 
\end{eqnarray}
by using the prescriptions for the toroidal magnetic 
field in Eqs.~\ref{eq:alf3} and \ref{eq:alf4}, respectively.  
In the following we refer to these models as 
WM3 and WM4, respectively.  
 
Equations~\ref{eq:xalfb}, \ref{eq:xalfc}, \ref{eq:xdiff}, and \ref{eq:xrec} 
give the ratio $x$ between the magnetospheric and the 
corotation radius (we assume $x\lesssim1$), for fixed values of 
$\xi$, $\gamma$, $\eta$, $\alpha$, $\mu_{30}$, $m$, $P_{\rm s}$, 
and $\dot{M}_{16}$.  
Some of these parameters are measured 
or constrained through observations ($\dot{M}$, $P_{\rm s}$, $\mu$, $m$);  
other parameters are still poorly determined by current theory: the values of 
$\xi$ and $\alpha$ (see Sect. \ref{sec:wmodel}) are uncertain by at least 
an order of magnitude \citep{king}, $\eta$ is in the 0.2-1 range 
and $\gamma$, $\gamma_{max}$ can be larger than 1 \citep{wang95}. 

In analogy to what we have done in Sect~\ref{sec:magngl}, we use here the definition of 
the critical fastness parameter and define $x_{\rm c}^{\rm w}$=($\omega_{\rm c}^{\rm w}$)$^{2/3}$.  
In this case, Eqs.~\ref{eq:xalfb}, \ref{eq:xalfc}, \ref{eq:xdiff}, and \ref{eq:xrec} translate into  
\begin{eqnarray}
&& x_{\rm w}^{-211/80}(1-x_{\rm w}^{3/2})^{1/2} = \nonumber \\
&& {x_{\rm c1}^{\rm w}}^{-211/80}(1-{x_{\rm c1}^{\rm w}}^{3/2})^{1/2}(\dot{M}_{16}/\dot{M}_{\rm tr16})^{4/5}, 
\label{eq:xalfbcri}
\end{eqnarray}
\begin{eqnarray}
&& x_{\rm w}^{-85/32}(1-x_{\rm w}^{3/2})^{1/2} = \nonumber \\
&& {x_{\rm c2}^{\rm w}}^{-85/32}(1-{x_{\rm c2}^{\rm w}}^{3/2})^{1/2}
(\dot{M}_{16}/\dot{M}_{\rm tr16})^{63/80},  
\label{eq:xalfccri}
\end{eqnarray}
\begin{equation}
x_{\rm w}^{-7/2}-x_{\rm w}^{-2}=({x_{\rm c3}^{\rm w}}^{-7/2}-{x_{\rm c3}^{\rm w}}^{-2})\dot{M}_{16}/\dot{M}_{\rm tr16},  
\label{eq:xdiffcri}
\end{equation}
and 
\begin{equation}
x_{\rm w}^{-7/2}-x_{\rm w}^{-2}=({x_{\rm c4}^{\rm w}}^{-7/2}-{x_{\rm c4}^{\rm w}}^{-2})\dot{M}_{16}/\dot{M}_{\rm tr16},   
\label{eq:xreccri}
\end{equation}
respectively. Here $x_{\rm c1}^{\rm w}$=0.966, $x_{\rm c2}^{\rm w}$=0.967, 
$x_{\rm c3}^{\rm w}$=0.915, $x_{\rm c4}^{\rm w}$=0.967, and we defined $\dot{M}_{\rm tr16}$ as 
the mass accretion rate (in unit of 10$^{16}$~g~s$^{-1}$) for which $x_{\rm w}$=$x_{\rm c}^{\rm w}$.  
In analogy to what we found for Eq.~\ref{eq:xglcri}, Eqs.~\ref{eq:xalfbcri}, \ref{eq:xalfccri}, 
\ref{eq:xdiffcri}, and \ref{eq:xreccri} show that all uncertain parameters cancel out and 
the magnetospheric radius can be easily estimated, as a function of the mass accretion rate $\dot{M}_{16}$, 
provided $\dot{M}_{\rm tr16}$ is somehow constrained by the observations. This is carried out in 
Sect.~\ref{sec:application} for the sample of X-ray powered pulsars we selected in the 
present study. 
\begin{figure}
\centering
\includegraphics[width=8.0 cm]{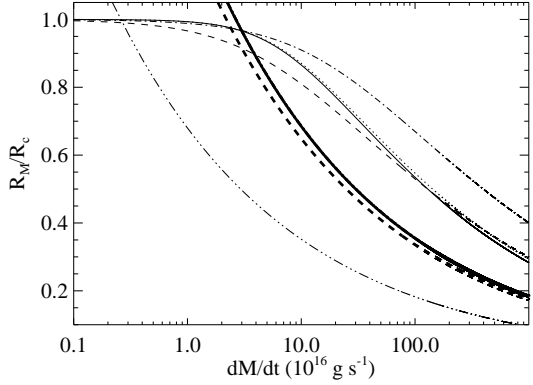}
\caption{The ratio $x$=$R_{\rm M}/R_{\rm co}$, as a function of the mass accretion rate, 
for different prescriptions of the toroidal magnetic fields and $\dot{M}_{\rm tr16}$=3.  
The solid line is for Eq.~\ref{eq:xalfbcri} (WM1), dotted line for Eq.~\ref{eq:xalfccri} (WM2),  
dashed line for Eq.~\ref{eq:xdiffcri} (WM3), and dot-dashed line for Eq.~\ref{eq:xreccri} (WM4). 
Thicker lines are the constant magnetic pitch approximations corresponding to each of the 
above prescriptions (note that the approximations corresponding to Eqs.~\ref{eq:xalfbcri}, 
\ref{eq:xalfccri}, and \ref{eq:xreccri} almost overlap). The triple-dot-dashed 
line represents the GLM (Eq.~\ref{eq:xglcri}).}
\label{fig:rffeff}
\end{figure}

A similar calculation can be applied to the case of the constant magnetic pitch approximation. 
We define $x_{\rm cmp}$=$R_{\rm M}^{\rm w}$/R$_{\rm co}$,  
and rewrite Eq.~\ref{eq:rmwapprox} as 
\begin{equation}
x_{\rm cmp}=4.175 \xi_{\rm p} \mu_{30}^{4/7} \dot{M}_{16}^{-2/7} M_1^{-10/21} P_{\rm s}^{-2/3}. 
\label{eq:xff}
\end{equation}
Using the definition of the critical fastness parameter, Eq.~\ref{eq:xff} translates into 
\begin{equation}
x_{\rm cmp}=x_{\rm c}^{\rm w} (\dot{M}_{16}/\dot{M}_{\rm tr16})^{-2/7},  
\label{eq:xffcri}
\end{equation}
where $x_{\rm c}^{\rm w}$=$x_{\rm c1}^{\rm w}$, $x_{\rm c2}^{\rm w}$, $x_{\rm c3}^{\rm w}$, 
$x_{\rm c4}^{\rm w}$, depending on the prescription used for the toroidal magnetic 
field. Note that, in the constant magnetic pitch approximation, 
the magnetospheric radius does not depend on the prescription used for the toroidal field:  
using a different equation for $B_{\phi}$ only affects the value of the critical fastness 
parameter that must be used in Eq.~\ref{eq:xffcri}. 
In Fig.~\ref{fig:rffeff} we arbitrarily fixed $\dot{M}_{\rm tr16}$=3 and compare the values 
of $x$=$R_{\rm M}$/$R_{\rm co}$, as a function of the mass accretion rate, obtained 
by solving numerically Eq.~\ref{eq:rmw} and using the constant pitch approximation 
for all the toroidal magnetic field prescriptions discussed in Sect.~\ref{sec:wmodel}. 
Despite all models approach the same asymptotic behaviour $x\propto\dot{M}^{-2/7}$ 
in the limit $x$$\ll$1, the constant pitch approximation results in a systematic smaller magnetospheric 
radius, for any considered value of the mass accretion rate and prescription 
of the toroidal magnetic field. 
On the other hand, in the limit $x$$\sim$1, the solution obtained with the constant 
magnetic pitch approximation significantly differs from the numerical solutions obtained with 
Eq.~\ref{eq:rmw}. In particular, the latter results in a magnetospheric radius that approaches 
the corotation radius more gradually, while decreasing the mass accretion rate: the range of $\dot{M}$ 
for which $x$=$R_{\rm M}/R_{\rm co}$$<$$x_{\rm c}^{\rm w}$ (and thus the NS spins down), is  
much larger with respect to the range obtained by assuming a constant magnetic pitch angle 
(see also Sect.~\ref{sec:wmodel}).  The discrepancy between these results demonstrate that 
the constant magnetic pitch approximation does not provide a reliable estimate of the 
magnetospheric radius in the WM. Therefore, we do not use Eq.~\ref{eq:xffcri} in the application 
to X-ray accretion powered pulsars in Sect.~\ref{sec:application}.  
For comparison we plot in Fig.~\ref{fig:rffeff} also Eq.~\ref{eq:xglcri}, which represents 
the magnetospheric radius in the GLM. The differences between this curve and those derived by using 
the constant magnetic pitch approximation and the WM are due to the different values  
of the critical fastness parameter in the GLM and WM 
(0.35 in the GLM and 0.875-0.95 in the WM; see Eqs.~\ref{eq:xglcri} and \ref{eq:xffcri}). 

\section{The beat frequency model}
\label{sec:BFM} 

Besides the threaded disk model, another probe of the position of 
the inner disk radius is offered by observations of QPOs 
in accreting NSs. These timing features \citep{klis95} have been 
detected in the X-ray flux of a number of astrophysical sources, especially old accreting NSs 
and black hole candidates in LMXBs but also in young accreting X-ray pulsars 
in high mass X-ray binaries. 
LMXBs often display a complex variety of simultaneous QPO modes, 
with frequencies ranging from few Hz up to  $\sim$1~kHz. 
On the contrary, young X-ray pulsars mostly display a single QPO, with 
a considerably lower frequency $\nu_{\rm QPO}$$\sim$0.008-0.2~Hz   
(see Table~\ref{tab:source} and references therein).  
Different models have been developed in order to interpret  
the nature of this X-ray variability.  
The fastest variability, manifested through 
kHz QPOs and timescales of $\sim$ms, must be generated by phenomena 
occurring in the innermost regions of the accretion disk and reflect 
the fundamental frequencies of motions in the close vicinity of the compact object 
\citep[see e.g.,][]{klis95}. 
On the contrary, mHz QPOs observed in accretion powered X-ray pulsars 
result from variability phenomena occurring farther away from 
the NS (the relevant timescales are of hundreds of seconds). 
In these sources, a magnetic field of order $\gtrsim$10$^{12}$~G 
disrupts the disk flow at the magnetospheric radius 
$R_{\rm M}$$\sim$10$^{8}$~cm, and thus the orbital motion at this radius 
provides an obvious source of variability. However, the involved time scales 
at $R_{\rm M}$ are a few tens of seconds at the most, and the beat  
between the orbital frequency at this radius and the spin frequency of the NS 
is generally invoked in order to interpret the observational properties of 
the slower (mHz) QPOs that are observed in these systems.  

According to beat frequency models \citep[BFM,][]{alpar,lamb85},
matter from inhomogeneities orbiting at the 
inner disk boundary ($R_{\rm M}$) is gradually removed 
through the interaction with the neutron star magnetosphere, thus giving 
giving rise to a modulation in the accretion rate and source luminosity.  
Therefore the QPO frequency $\nu_{\rm QPO}$ results from the beat between the 
orbital frequency $\nu_{\rm orb}$ of the blobs at $R_{\rm M}$ and the spin frequency 
of the NS $\nu_{\rm s}$=2$\pi$$\Omega_{\rm s}$, {\it i.e.} 
$\nu_{\rm QPO}$=$\nu_{\rm orb}$($R_{\rm M}$)-$\nu_{\rm s}$. In practice  
$\nu_{\rm orb}$($R_{\rm M}$) is well approximated with $\nu_{\rm K}$($R_{\rm M}$), 
{\it i.e.} the Keplerian frequency at $R_{\rm M}$ (see below) and the 
above equation can be solved for the magnetospheric radius. This gives:         
\begin{equation}
R_{\rm M}^{\rm BFM}=3.3\times10^8 \left(\frac{0.3 Hz}
{\nu_{\rm s}+\nu_{\rm QPO}}\right)^{2/3}
\left(\frac{M}{M_{\odot}}\right)^{1/3} ~{\rm cm}, 
\label{eq:rqpo}  
\end{equation}
or 
\begin{equation}
x_{\rm BFM}=2.2 \left(\frac{0.3 Hz}{\nu_{\rm s}+\nu_{\rm QPO}}\right)^{2/3} 
P_{\rm s}^{-2/3}, 
\label{eq:xqpo} 
\end{equation}
where $x_{\rm BFM}$=$R_{\rm M}^{\rm BFM}$/$R_{\rm co}$.   
By assuming that the BFM applies, QPOs in X-ray pulsars can be used to probe 
the physical condition of the disk flow at the inner disk radius; in particular,  
Eq.~\ref{eq:xqpo} allows for a straightforward estimate of $R_{\rm M}$. 
We note that the magnetospheric radius is, by definition,  
the innermost radius at which the disk plasma maintains a nearly Keplerian orbit. 
It might be expected that the beat between the disk inhomogeneities (blobs) and 
the magnetosphere takes place when the former have achieved a sub-Keplerian orbital 
frequency. However, in this case a blob would not be centrifugally supported any longer
and would drift inwards, where the magnetic stresses 
would rapidly bring the blob 
into corotation with the NS; therefore modulated accretion at the 
beat frequency could not extend over several beat cycles, as required 
to explain the QPO Q-factors \citep[which 
range from a few to tens in most cases; see e.g.][]{klis04}. 
Therefore, it can be ruled out that disk inhomogeneities giving rise to modulated accretion
at the beat frequency orbit at substantially slower frequencies than Keplerian.  
On the other hand, a larger orbital frequency at $R_{\rm M}$  than the corresponding 
Keplerian frequency, can be certainly ruled out by the effect of viscosity 
in the accretion disk.   

\section{Applications to accretion powered X-ray pulsars} 
\label{sec:application}
Here we apply the calculations discussed in the previous 
sections to accretion powered X-ray sources. In particular, 
we selected the sources which displayed QPOs in their X-ray flux, 
as well as evidence for transitions between spin-up and spin-down states.   
For each source, we use the luminosity measured when QPO were detected ($L_{\rm QPO}$)
and the luminosity at which spin-up/spin-down transitions took place ($L_{\rm tr}$) 
in order to estimate the magnetospheric radius within the BFM (see Sect.~\ref{sec:BFM})
and magnetically threaded disk models (see Sect.~\ref{sec:magngl} and \ref{sec:magnet}), 
respectively. A comparison between these estimates of the magnetospheric radius is then carried out.  
In Table \ref{tab:source} we report the relevant values of $L_{\rm QPO}$ and $L_{\rm tr}$ we used, 
while in appendix~\ref{sec:observations} we give a brief summary of the properties of each 
source in our sample. 

In order to calculate the magnetospheric radius in the threaded disk models, 
we first use the observations of spin-up/spin-down transitions. 
According to the threaded disk models, these transitions 
are the results of changes in the sign (from positive to negative)  
of the torque $n$ acting onto the NS. Therefore, the luminosity $L_{\rm tr}$ can be used  
to constrain the value of $\dot{M}_{\rm tr}$, at which, according to the models, the torque $n$ 
is expected to undergo a sign reversal (see Eqs.~\ref{eq:gltorque} and \ref{eq:wtorque}).  
In accretion powered X-ray pulsars, the conversion between $L_{\rm tr}$ and $\dot{M}_{\rm tr}$ can 
be obtained by using the relation
\begin{equation}
L_{\rm tr36}=1.3 \zeta \dot{M}_{tr16}(M/M_{\odot})(R_{\rm NS}/10^6), 
\label{eq:lx}
\end{equation}
where $L_{\rm tr36}$ is the X-ray luminosity in unit of 10$^{36}$~erg~s$^{-1}$   
and $\zeta$$\sim$1 is an efficiency factor that takes into account, e.g. geometrical 
and bolometric corrections (see later in this section).  
\begin{figure*}
\centering
\includegraphics[width=16.0 cm]{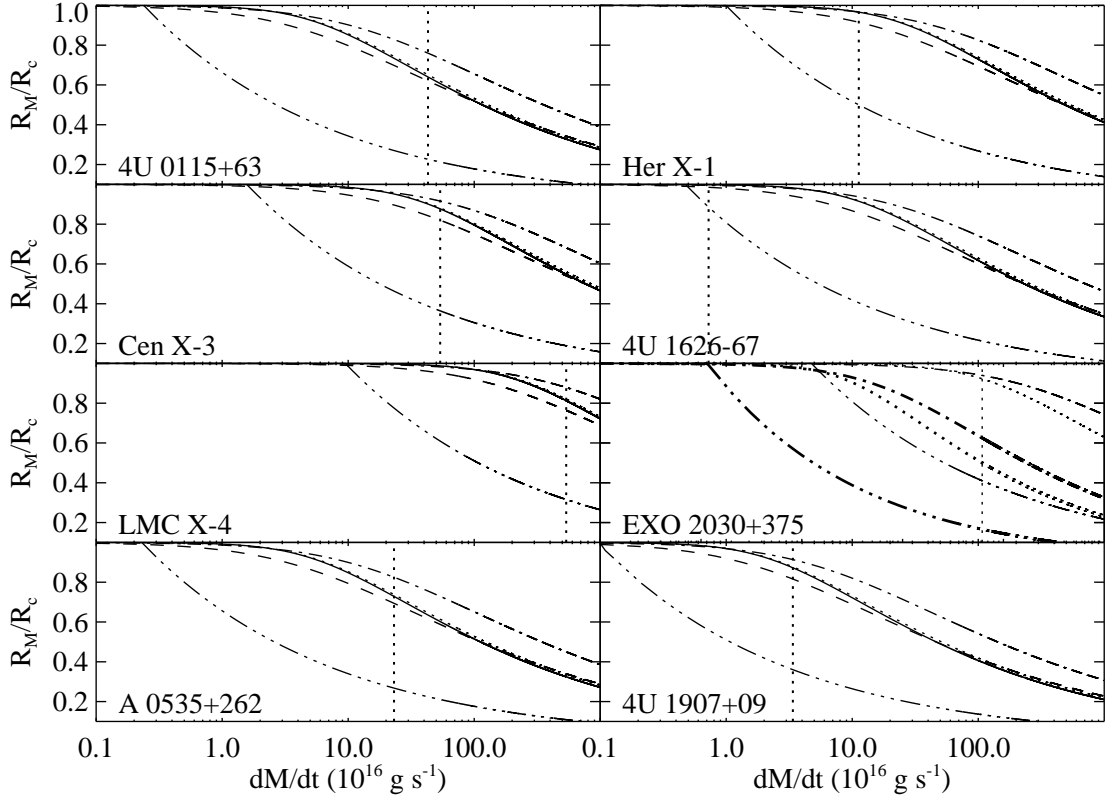}
\caption{The ratio between the magnetospheric radius and the corotation radius, 
as a function of the mass accretion rate, for sources in Table~\ref{tab:source}. 
The solid line is for Eq.~\ref{eq:xalfbcri} (WM1), dotted line for Eq.~\ref{eq:xalfccri} (WM2),  
dashed line for Eq.~\ref{eq:xdiffcri} (WM3), dot-dashed line for Eq.~\ref{eq:xreccri} (WM4), and 
triple-dot-dashed line for Eq.~\ref{eq:xglcri} (GLM). In the case of EXO\,2030+375, 
spin-up/spin-down transitions were observed at luminosities in the range 
$\sim$10$^{38}$-2.4$\times$10$^{36}$~erg~s$^{-1}$, and we represented in the figure 
only those models that provide the higher and lower bound of $R_{\rm M}/R_{\rm co}$ 
as a function of the mass accretion rate. The thinner lines are for 
$L_{\rm tr}$=10$^{38}$~erg~s$^{-1}$, whereas the thicker lines are for 
$L_{\rm tr}$=2.4$\times$10$^{36}$~erg~s$^{-1}$. 
In all panels the dotted vertical lines represent values of the mass accretion rate corresponding  
to the luminosity $L_{\rm QPO}$ at which QPOs are observed in each source of our sample.}  
\label{fig:total}
\end{figure*}

Once $\dot{M}_{tr16}$ is determined, the magnetospheric radius in both the 
GLM and WM can be easily estimated as a 
function of the mass accretion rate, since all the uncertain parameters cancel out  
(see Sect.~\ref{sec:magngl} and \ref{sec:magnet}). 
This is shown in Fig.~\ref{fig:total} for the X-ray pulsars 
in our sample (we assumed a NS mass of $m$=1.4 and a radius of $R_{\rm NS}$=10$^{6}$~cm). 
For each source we plot in the panels of this figure the derived values of the magnetospheric radius 
(units of the corotation radius), as a function of the mass accretion rate 
(units of 10$^{16}$~g~s$^{-1}$), for the threaded disk models described by 
Eq.~\ref{eq:xglcri} (GLM, triple-dot-dashed line), Eq.~\ref{eq:xalfbcri} (WM1, solid line), 
Eq.~\ref{eq:xalfccri} (WM2, dotted line), Eq.~\ref{eq:xdiffcri} (WM3, dashed line), and 
Eq.~\ref{eq:xreccri} (WM4, dot-dashed line). 
At this point we use Eq.~\ref{eq:lx} and values of $L_{\rm QPO}$ in Table~\ref{tab:source} 
to calculate the mass accretion rate at which QPOs are observed in each X-ray pulsar of our sample.  
The derived mass accretion rates are represented in panels of Fig.~\ref{fig:total} with dotted 
vertical lines. 
For each source, the intersection between the dotted vertical line and the curves representing the 
magnetic threaded disk models gives the magnetospheric radius predicted by these models at the 
mass accretion rate corresponding to the $L_{\rm QPO}$ luminosity.  
In particular, the intersection with the curves that represent Eq.~\ref{eq:xalfbcri},  
\ref{eq:xalfccri}, \ref{eq:xdiffcri}, and \ref{eq:xreccri} give the range 
of allowed values of $x_{\rm w}$=$R_{\rm M}^{\rm w}$/$R_{\rm co}$, {\it i.e.} the magnetospheric radius in the WM 
(in unit of the corotation radius) calculated at the mass accretion rate that corresponds 
to $L_{\rm QPO}$. We indicate this parameter with $x_{\rm QPO}^{\rm w}$ in Table~\ref{tab:source}. 
Similarly, $x_{\rm QPO}^{\rm gl}$ is the value of $x_{\rm gl}$=$R_{\rm M}^{\rm gl}$/$R_{\rm co}$ 
for which the vertical dotted line intersects the curve from the GLM (Eq.~\ref{eq:xglcri}).  
Finally, we derive for each source the value of the magnetospheric radius in the BFM 
at the mass accretion rate that corresponds to  $L_{\rm QPO}$ by using Eq.~\ref{eq:xqpo}. 
We call this parameter $x_{\rm QPO}^{\rm BFM}$ 
(a range of values for $x_{\rm QPO}^{\rm BFM}$ is indicated in Table~\ref{tab:source} 
only for those sources which displayed more than one QPO frequency). 

With values of $x_{\rm QPO}^{\rm w}$, $x_{\rm QPO}^{\rm gl}$, and 
$x_{\rm QPO}^{\rm BFM}$ at hand, the GLM and WM can be tested 
against observations of accretion powered X-ray pulsars. 
Looking at values of these parameters in Table~\ref{tab:source}, we note that the selected 
sample of sources can be roughly divided into two groups. The first 4 sources 
(\her,\ \0115,\ \cen,\ \lmc\ ) displayed 
QPO frequencies that, if interpreted in terms of the BFM, agree with predictions  
of the WM. In fact, in these cases, $x_{\rm QPO}^{\rm BFM}$ and $x_{\rm QPO}^{\rm w}$ 
have similar values, whereas $x_{\rm QPO}^{\rm gl}$ is typically a factor of 2-3 smaller 
(for \0115\ and \lmc\ the small discrepancy between $x_{\rm QPO}^{\rm BFM}$ and 
$x_{\rm QPO}^{\rm w}$ can be easily accounted for, e.g. by assuming small bolometric corrections in the 
X-ray luminosities $L_{\rm QPO}$ and $L_{\rm tr}$, see Sect.~\ref{sec:application}). 
Values of $x$ close to $\sim$1, as measured for these four sources, imply  
a magnetospheric radius that is very close to the corotation radius for the luminosities at which QPOs  
are detected (for example, in the cases of \her\ and \lmc,\ 
the magnetospheric and corotation radii differ by less than few percent); therefore,   
in the following we refer to these sources as ``fast rotators''.   
Instead, results obtained for \exo,\ \a0535,\ and \4u1907\ suggest the GLM 
is better suited to account for observations of this second group of sources. 
In this case values of $x_{\rm QPO}^{\rm BFM}$ are much closer to $x_{\rm QPO}^{\rm gl}$ 
than $x_{\rm QPO}^{\rm w}$. 
However, only for \a0535\ a good agreement between the BFM and the GLM is obtained. 
In the other two cases (\exo\ and \4u1907\ ) $x_{\rm QPO}^{\rm gl}$ is at least a factor of 
2 larger than $x_{\rm QPO}^{\rm BFM}$ ($x_{\rm QPO}^{\rm w}$ is a factor of 2-3 larger than $x_{\rm QPO}^{\rm gl}$). 
In these sources the magnetospheric radius at the mass accretion rate corresponding to $L_{\rm QPO}$ is well inside the 
corotation radius ($x$$\ll$1), and thus in the following we refer to them as ``slow rotators''.  
The QPO properties of \1626\ suggest a magnetospheric radius close to the corotation radius 
($x_{\rm QPO}^{\rm BFM}$=0.83), like in the case of fast rotators, but they are well interpreted within the GLM. 
This source might thus be a sort of ``transition object'' between fast and slow rotators. 

The conversion in Eq.~\ref{eq:lx} between observed X-ray luminosity and mass accretion rate is affected 
by several uncertainties. Besides the NS mass and radius, 
effects which can make $\zeta$ in Eq.~\ref{eq:lx} differ from unity,  
such as non-isotropic emission and bolometric corrections, should be kept in mind. 
Despite these uncertainties, we note that the results derived in this and the next 
section are virtually insensitive to variations by a factor of few in $L_{\rm tr}$ and $L_{\rm QPO}$. 
This is due to the weak dependence of the magnetospheric radius on the mass accretion rate.  
In all models discussed in Sect.~\ref{sec:review}, the steepest dependence of $R_{\rm M}$ on $\dot{M}$ 
is $\propto$$\dot{M}^{-2/7}$; therefore, an uncertainty by a factor 2-3 in the X-ray luminosity 
(and thus on $\dot{M}$, see Eq.~\ref{eq:lx}) would cause a 20-30\% change in the magnetospheric 
radius at the most.

\subsection{Two case studies: \1626\ and \cen\ }
\label{sec:casestudies}

\begin{figure}
\centering
\includegraphics[scale=0.47]{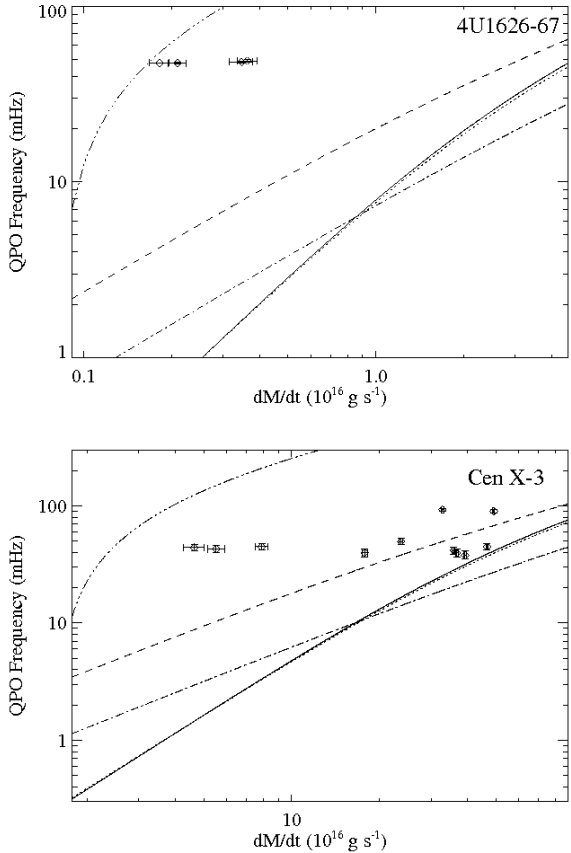}
\caption{Predicted QPO frequencies in the BFM when the magnetospheric radius 
is calculated according to Eq.~\ref{eq:xalfbcri} (WM1, solid line), Eq.~\ref{eq:xalfccri} (WM2, dashed line)   
Eq.~\ref{eq:xdiffcri} (WM3, dotted line), Eq.~\ref{eq:xreccri} (WM4, dash-dotted line), and Eq.~\ref{eq:xglcri} 
(GLM, triple-dot-dashed line). The upper panel is for \1626,\ whereas the lower panel is for 
\cen.\ In both cases QPO frequencies inferred from observations of these sources at different 
X-ray fluxes are shown together with measurement errors. These data taken from \citet{kaur08}, 
\citet{krauss07}, and \citet{raichur}; we used Eq.~\ref{eq:lx} to convert 
between X-ray fluxes and mass accretion rates.}
\label{fig:casestudies}
\end{figure}
In this section we discuss further the cases of \1626\ and \cen,\ for which detailed 
studies of the long term variations of the QPO frequency with the X-ray flux 
were recently published. 

In particular, \citet{kaur08} studied QPOs in \1626\ at different X-ray fluxes and compared the observed 
frequencies with those calculated by using the BFM and with the magnetospheric 
radius determined based on the GLM. 
These authors noted a discrepancy between the observations and the predictions, and argued that 
the BFM might not apply to this source. 
In Fig~\ref{fig:casestudies} (upper panel) we show the same calculation, but 
included also the QPO frequencies estimated by using the BFM with the magnetospheric radius as determined in the WM.  
In this plot the solid line corresponds to the QPO frequency predicted by the BFM when   
Eq.~\ref{eq:xalfbcri} (WM1) is assumed for the magnetospheric radius, the dashed line is for  
Eq.~\ref{eq:xalfccri} (WM2), dotted line for Eq.~\ref{eq:xdiffcri} (WM3), and the dash-dotted line for 
Eq.~\ref{eq:xreccri} (WM4). The triple-dot-dashed line represents the QPO frequencies 
predicted by the BFM when the magnetospheric radius is calculated according to the GLM (Eq.~\ref{eq:xglcri}). 
We selected those data from \citet{kaur08} and \citet{krauss07} for which QPO frequencies and X-ray fluxes 
were measured simultaneously, and used Eq.~\ref{eq:lx} to convert these fluxes into mass accretion rates.  
In the upper panel of Fig.~\ref{fig:casestudies} we show that, due to the different prescriptions available for the  
magnetospheric radius ({\it i.e.} the GLM and the WM), the region of predicted QPO frequencies in the BFM, 
as a function of the mass accretion rate, is very broad and all  
observational measurements lie within this region.  

The lower panel of Fig.~\ref{fig:casestudies} show the case of \cen.\ 
We used data from \citet{raichur}. 
In this work, the authors showed that the QPO frequency of this source 
has almost no dependence on the X-ray flux. By using the GLM to calculate the magnetospheric 
radius, they argued that, if the BFM applies, then the long term X-ray 
intensity variations of \cen\ are likely due to obscuration by an aperiodically precessing warped disk,  
rather than being related to changes in the mass accretion rate (and thus location of the inner disk radius). 
In fact, in the latter case the QPO frequency would be expected to vary 
according to Eqs.~\ref{eq:xqpo} and \ref{eq:rmgl}.   
However, our calculations show  that all measured QPO frequencies  
lie inside the region spanned by different magnetically threaded disk models. 
We conclude that the observations of \1626\ and \cen\ do not support simple applications 
of either the GLM or WM to the BFM. We further comment on this in Sect.~\ref{sec:discussion}.

\section{Discussion}
\label{sec:discussion} 

The results obtained in the previous section indicate neither the GLM nor the 
WM, when used in conjunction with the BFM, are able to reproduce the range of observations 
discussed here for the entire sample of X-ray pulsars. 
We also note that for all sources in Fig.~\ref{fig:total}, the magnetospheric radius 
in the GLM turns out to be somewhat smaller than that derived by using the WM. 
This point was discussed also by \citet{wang96}, who   
suggested that the reason for this disagreement resides in the different prescription of 
the toroidal field used in the two models: the assumed $B_{\rm\phi}$ in the GLM 
implies a larger magnetic torque that spins down the NS more efficiently 
and reduces the value of the critical fastness parameter. As a consequence, the GLM magnetospheric 
radius is located closer to the NS (see Fig.~\ref{fig:total}). 
We showed here that the magnetospheric radius predicted by the GLM is still too large to account 
for observations of the QPOs over the entire sample of the slow rotators (see Table~\ref{tab:source}).
This remains true even when the GLM is revised to include a more accurate prescription 
of the toroidal magnetic field, which leads to larger values of the critical fastness parameter 
\citep{wang87,ghoshlamb92,ghoshlamb95}.  
 
At odds with the GLM, the magnetospheric radius in the WM approaches 
the corotation radius more gradually as the mass accretion rate decreases, a  
result that seems to account for observations of fast rotating sources (see Table~\ref{tab:source}).
However, in the cases of \1626\ and \cen,\ for which detailed studies of the long term variations 
of the QPO frequency with the X-ray flux are available, the WM is not able to reproduce the observations.   
It was also noted that the treatment of the NS poloidal field screening 
by currents flowing onto the disk surface in this model might be oversimplified 
\citep{ghoshlamb92,ghoshlamb95}. 

Furthermore, in Sect.~\ref{sec:wmodel} we pointed out that an important caveat in the WM is that the interaction 
between the accretion disk and the NS magnetic field takes place in a similar fashion over the whole 
accretion disk. This is at odds with the GLM that predicts the strong coupling between 
the NS and the disk takes place mostly within a small boundary layer, such that this region alone 
determines the position where the disk terminates ({\it i.e.} $R_{\rm M}$). 
On the other hand, the theory of the boundary layer envisaged in the GLM might 
not be applicable to fast rotators, being the radial extent of the 
boundary layer of the same order of the separation between the magnetospheric and 
the corotation radii in these cases\footnote{Note that \citet{ghoshlamb2} showed steady state 
accretion might not be allowed in their model for $\omega_{\rm c}$$\gtrsim$0.95.}. 
Some works have investigated the importance of the boundary layer in the threaded disk model 
\citep{liwang2, liwickra, liwang, alpar2}\footnote{``Torqueless accretion'' 
\citep{liwickra2, liwickra} was not considered here since that mechanism is unlikely to 
be applicable to accretion powered X-ray pulsars \citep{wang97b, romanova}.}.  
\citet{liwang2} suggested that there exists an uncertainty by a factor of 4 in the \citet{wang87} 
equation defining the magnetospheric radius; using this result,    
\citet{liwang} demonstrated that the boundary layer in NS accreting binaries 
can survive the destructing action of the NS magnetic field down to a radius 
$\simeq$0.8$R_{\rm M}$. The boundary layer might thus be significantly larger than previously 
thought \citep{ghoshlamb1}. However, the derived corrected value of the critical fastness 
parameter ($\sim$0.71-0.95) does not differ much from previous estimates \citep{wang95} and 
the problem of slow rotating sources remains open.     
Similar results were obtained by \citet{alpar2}, which demonstrated that the width of the boundary 
layer might be a strong function of the fastness parameter: they found that broad ``boundary layers'' 
are expected for spinning-up sources, whereas much reduced boundary layers should be expected for 
sources in a spin-down state (these boundary layers are typically a factor $\sim$6-60 less wider  
than those found for spinning-up sources). 
However, a general analytical equation for the magnetospheric radius 
cannot be easily derived, due to the presence of few additional parameters in their model.  
Broad boundary layers were found also in the simulations by \citet{romanova}. These authors 
found a reasonable agreement with the predictions of the GLM, with the inner region 
of the disk behaving like a boundary layer, while the outer region is only partially coupled with the magnetic 
field of the star. 
These numerical simulations suggested a critical fastness parameter of $\sim$0.6. 
Despite this value is in between the values obtained within the GLM and the WM, 
it cannot account for observations of both fast and slow rotators. Our results in Table~\ref{tab:source} 
imply that the critical fastness parameter cannot be constant for all these sources. 
A more general solution for the magnetic threaded disk model might be found in the future 
in which the WM and the GLM give the limiting cases of fast and slow 
rotation, respectively.  

The present study suggests that all the discussed limitations of both the WM and GLM might be 
the reason why none of these models is able to reproduce the combination of QPO and 
torque behaviour observed at different X-ray luminosity levels in the X-ray pulsars considered. 
Alternatively, the BFM might not be applicable to (all) QPOs observed from X-ray pulsars.

\section{Conclusions}
\label{sec:conclusions}
We showed that, if the BFM applies to the QPOs of X-ray pulsars, then 
the GLM and WM cannot completely account for observations 
of the sources in our sample.   
Instead, taking into accounts results in Table~\ref{tab:source}, we noted that these  
sources can be divided into two classes:
\begin{itemize}
\item Fast rotators, for which the Keplerian velocity of matter at the inner disk radius, as 
inferred from the application of the BFM to the observed QPO frequency, is close to the 
rotational velocity of the star. 
In this case, the magnetospheric radius inferred from the WM and the BFM predict 
QPO frequencies which seem in good agreement with the observations.   
However, we showed in Sect~\ref{sec:casestudies} that at least in the cases of \1626\ and \cen,\ 
the WM is not able to reproduce observations of QPO frequencies at different X-ray fluxes.  
\item Slow rotators, for which the Keplerian velocity of matter at the inner disk radius, as inferred from 
the application of the BFM to the observed QPO frequency, is well above the spin velocity of the NS. 
In this case, the magnetospheric radius derived from the BFM is less discordant with   
the predictions of the GLM. In fact, only for \a0535\ a good agreement between the GLM and 
the observations is obtained.  
For slow rotators the WM give a magnetospheric radius that is at least $\gtrsim$2-8 times larger 
than that derived from the BFM.  
\end{itemize} 

We conclude that either a more advanced theory of magnetically threaded disks is required, 
or that the BFM does not apply to (all) QPOs observed from X-ray pulsars.

\begin{appendix}
\section{Calculation of the torque for region ``C'' of the SS73 accretion disk model}
\label{app:A}
In order to evaluate $\rho$ and $h$ in Eqs. \ref{eq:alf2}  and \ref{eq:wtorque}, 
\citet{wang87} considered only the ``B'' region of the SS73  
accretion disk, {\it i.e.} the gas-pressure dominated region where electron scattering gives 
the main contribution to the opacity.  
Here we carry out the same calculation by using the expressions of 
$\rho$ and $h$ that are appropriate to region ``C'' of the SS73 accretion disk model 
(where the main contribution to the opacity is provided by free-free absorption). 
According to \citet{vietri}, the thermal pressure of disk matter has 
a radial dependence $p=p_{\rm M} (R_{\rm M}/R)^{21/8}$, where the subscript 
$M$ denotes quantities evaluated at the inner disk radius. 
From Eq.~\ref{eq:alf2} we get 
\begin{eqnarray}
\left\vert B_{\phi}(R)\right\vert & = & B_{\phi 0}\frac{\omega_{\rm s}^{23/16}}
{\vert1-\omega_{\rm s}\vert^{1/2}}\left(\frac{R_{\rm co}}{R}\right)^{69/32}\cdot \\ \nonumber 
&& \cdot\left\vert1-\left(\frac{R}{R_{\rm co}}\right)^{3/2}\right\vert^{1/2}, 
\label{ap1}
\end{eqnarray}
where the same notation as that of Sect.~\ref{sec:review} is used. 
Setting $y=(R/R_{\rm co})^{3/2}$ and using Eq.~\ref{ap1} into Eq.~\ref{eq:wtorque} 
we obtain 
\begin{eqnarray}
n(\omega_{\rm s})&=&1+\frac{1}{3}\frac{\omega_{\rm s}^{23/16}}{\vert1-\omega_{\rm s}
\vert^{1/2}}\cdot \\ \nonumber  
&& \cdot\left[\int_{\omega}^1\frac{(1-y)^{1/2}}{y^{39/16}}dy 
-\int_{1}^{\infty}\frac{(y-1)^{1/2}}{y^{39/16}} dy\right]. 
\end{eqnarray}
By numerically evaluating integrals in the above equation 
we find $\omega_{c}=0.95$. 

\section{Values of $L_{\rm tr}$ and $L_{\rm QPO}$ for accretion 
powered X-ray sources}
\label{sec:observations}
Here we briefly summarize the relevant observations of the accretion powered X-ray pulsars considered in 
Table~\ref{tab:source}, in order to explain values used for the luminosities 
$L_{\rm tr}$ and $L_{\rm QPO}$. 
\begin{itemize}
\item {\em Her\,X-1}: Her\,X-1 is one of the best studied X-ray binary system. It consists of a 
$\sim$1.24~s spinning NS and a A/F companion (the orbital period is 1.7~day). The X-ray flux 
displays a regular modulation at a 35~day period, that has been associated to 
the precession of a highly warped accretion disk that periodically obscures the NS 
\citep{petterson75,choi94,wilson97,dalfiume98,par99,klochkov07}. 
This suggested that transitions between high (``main-on states'') 
and low luminosity states (the ``anomalous'' low states) of this source  
are to be interpreted as due to local obscuration phenomena, rather than  
large changes in the mass accretion rate.  
Evidence in favor of this interpretation has been recently obtained through 
detailed phase-resolved spectroscopy \citep{zane04} as well as observations of X-ray heating of the companion 
star \citep{boroson00}. Since QPOs and spin-up/spin-down transitions were observed during both 
high and low luminosity states, for the purpose of this paper we assume 
$L_{\rm tr}$=$L_{\rm QPO}$=2.1$\times$10$^{37}$~erg~s$^{-1}$ (a distance of 5~kpc is considered),   
where the latter is the typical main-on state luminosity. 

\item{\em 4U\,0115+63}: 4U\,0115+63 is a binary system hosting a 3.6~s spinning 
NS orbiting a Be companion (the orbital period is 24.3~day). The distance to the source 
is $\sim$8~kpc \citep{negueruela01}. 
An HEAO observation caught this source in outburst \citep[the typical outburst luminosity is 
$\sim$8$\times$10$^{37}$~erg~s$^{-1}$,][]{camp01}, 
and a prominent peak in the power spectrum of the X-ray light curve was detected at 62~mHz \citep{soong89}. 
\citet{tam} reported that the spin-up trend the source usually displayed while in outbursts reversed 
during lower luminosity states. These typically occurred at 5$\times$10$^{35}$~erg~s$^{-1}$ \citep{camp01}.  
  
\item{\em Cen\,X-3}: Cen\,X-3 is a high mass X-ray binary with a spin period of $\sim$4.8~s and an 
orbital period of $\sim$2.1~day. The companion star is an O-type supergiant and the distance is 
estimated to be $\sim$8~kpc \citep[][and references therein]{burderi00}. 
The presence of a $\sim$35~mHz QPO in the power spectrum of this source was first discovered by 
\citet{take}, after the source egress from an X-ray eclipse.    
The typical X-ray luminosity was determined with BeppoSAX, and is  
of order $\sim$1.0$\times$10$^{38}$~erg~s$^{-1}$ (0.12-100~keV).  
Cen\,X-3 has a secular spin-down trend \citep{bild}, but episodes of spin reversal 
were found to occur at a luminosity that is typically a factor of $\sim$3 below 
that of the post-eclipse high luminosity state \citep{howe}. 
We thus considered in Table~\ref{tab:source} that $L_{\rm QPO}$ is equal to the typical 
X-ray luminosity observed in the post-eclipse egress state, whereas 
$L_{\rm tr}$$\sim$1/3$L_{\rm QPO}$. 

\item{\em LMC\,X-4}: LMC\,X-4 is an accretion powered X-ray pulsar with a spin 
period of $\sim$13.5~s and an orbital period of $\sim$1.4~day. The X-ray luminosity 
of the system varies with a periodicity of $\sim$30.3~day, alternating between  
high ($\sim$2$\times$10$^{38}$~erg~s$^{-1}$) and low (a factor of $\sim$60 below) 
luminosity states \citep[for an estimated distance of $\sim$50~kpc,][]{woo}. 
This periodicity have been attributed to the effect of an obscuring tilted accretion disk 
\citep[][and references therein]{moon01}. 
During the high states, spin torque reversals were repeatedly observed, whereas  
during episodes of very bright flares ($\sim$10$^{39}$~erg~s$^{-1}$, 2-25 keV) QPOs were detected 
at frequencies in the $\sim$0.65-1.35~mHz range \citep[these were interpreted within  
the BFM by][]{moon01}. 

\item{\em 4U\,1626-67}: 4U\,1626-67 is a low mass X-ray binary hosting a $\sim$7.7~s spinning 
neutron star in a $\sim$42~min orbit around a $\sim$0.004~M$_{\odot}$ companion star. 
QPOs were detected in the X-ray observations of this source more than once 
\citep[for a review see, e.g.][]{kaur08}. A torque reversal was observed by 
\citet{cha97}, who also estimated a source distance of $\sim$3~kpc. 
For a review of the flux history of 4U\,1626-67, we reefer the reader to 
\citet{krauss07}. 

\item{\em EXO\,2030+375}: EXO\,2030+375 is a Be X-ray transient with an orbital period 
of $\sim$46~day, hosting a $\sim$42~s spinning neutron star, and located at a distance  
of 7.1~kpc \citep{wilson02}. 
In 1985 this source underwent a bright outburst 
(peak luminosity $\sim$2$\times$10$^{38}$~erg~s$^{-1}$), and a QPO at $\sim$213~mHz 
was detected \citep{angelini}.  Spin-up/spin-down transitions 
were observed more than once, at luminosities in the 
$\sim$10$^{38}$-2.4$\times$10$^{36}$~erg~s$^{-1}$ (1-20~keV) range. 

\item{\em A0535+262}: A0535+262 is a $\sim$103~s X-ray pulsar, orbiting 
a O9.7 companion star (the orbital period is $\sim$111~day). 
QPOs and spin reversals were best observed 
during the giant outburst in 1994 \citep{fin96}. This outburst was detected 
with BATSE in the energy 20-100~keV range, and the flux at the peak of the outburst 
was $\sim$6~Crab. Observations of this outburst at lower energies ($<$20~keV) were not available, 
but, based on previous results, \citet{fin96} estimated the 2-10 keV flux might not 
be larger than 2~Crab. Due to the uncertainties in this estimate we have not corrected  
values reported in Table~\ref{tab:source} for this factor. As discussed 
in Sect.~\ref{sec:application} an uncertainty of $\sim$30\% on the luminosity used to 
derive the position of the magnetospheric radius cannot affect much our results. 
Note that the distance used by \citet{fin96} to convert the observed flux into an 
X-ray luminosity is 2~kpc \citep{steele98}. 

\item{\em 4U\,1907+09}: 4U\,1907+09 is a $\sim$440~s spinning NS in a $\sim$8~day  
orbit around a supergiant companion. QPOs were discovered  
during an hour long flare at $\sim$6.3$\times$10$^{36}$~erg~s$^{-1}$ 
\citep[2-60~keV, and an assumed distance of 5~kpc,][]{zand98,cox05}. After 
the discovery of the pulsations by \citet{makishima92}, the 
NS in 4U\,1907+09 exhibited a steady spin-down for about 20~yrs. This trend changed in 
2006, when a torque reversal was observed at $\sim$2$\times$10$^{36}$~erg~s$^{-1}$ 
\citep[1-15~keV,][]{fritz}. 

\end{itemize}

\end{appendix}

\section*{Acknowledgments}
EB thanks University of Colorado at Boulder and JILA  
for hospitality during part of this work, and M. Falanga 
for useful comments. PG thanks Osservatorio Astronomico 
di Roma and University of Rome ``Tor Vergata'' for warm 
hospitality while part of this work was done. 
This work was partially supported through ASI 
and MIUR grants.


\begin{thebibliography}{}
\bibitem[\protect\citeauthoryear{Alpar et al.}{1985}]{alpar} 
Alpar, M.A., Shaham, J. 1985, Nature, 316, 239 

\bibitem[\protect\citeauthoryear{Angelini et al.}{1989}]{angelini} 
Angelini, L., Stella, L., Parmar, A.N. 1989, ApJ, 346, 906	

\bibitem[\protect\citeauthoryear{Bildsten et al.}{1997}]{bild} 
Bildsten, Lars, Chakrabarty, D., Chiu, J., et al. 1997, ApJs, 113, 367

\bibitem[\protect\citeauthoryear{Boroson et al.}{2000}]{boroson00}
Boroson, B., O'Brien, K., Horne, K., Kallman, T., Still, M.,  
Boyd, P.T., Quaintrell, H., Vrtilek, S.D. 2000, ApJ, 545, 399

\bibitem[\protect\citeauthoryear{Burderi et al.}{2000}]{burderi00}
Burderi, L., Di Salvo, T., Robba, N.R., La Barbera, A., Guainazzi, M. 2000, ApJ, 530, 429

\bibitem[\protect\citeauthoryear{Campana et al.} {2001}]{camp01} 
Campana, S., Gastaldello, F., Stella, L., et al. 2001, ApJ, 561, 924

\bibitem[\protect\citeauthoryear{Chakrabarty et al.}{1997}]{cha97} 
Chakrabarty, D., Bildsten, L., Grunsfeld, J.M. 1997 ApJ, 474, 414

\bibitem[\protect\citeauthoryear{Chakrabarty}{1998}]{cha} 
Chakrabarty, D. 1998, ApJ, 492, 342 

\bibitem[\protect\citeauthoryear{Choi et al.}{1994}]{choi94} 
Choi, C.S., Nagase, F., Makino, F., Dotani, T., Min, K.W. 1994, ApJ, 422, 799 

\bibitem[\protect\citeauthoryear{Cox et al.}{2005}]{cox05} 
Cox, N.L.J., Kaper, L., Foing, B.H., Ehrenfreund, P. 2005, A\&A, 438, 187 

\bibitem[\protect\citeauthoryear{Dal Fiume et al.}{1998}]{dalfiume98} 
dal Fiume, D., Orlandini, M., Cusumano, G., del Sordo, S., Feroci, M., 
Frontera, F., Oosterbroek, T., Palazzi, E., Parmar, A.N., Santangelo, A., 
Segreto, A. 1998, A\&A, 329 L41 

\bibitem[\protect\citeauthoryear{Erkut \& Alpar}{2004}]{alpar2} 
Erkut, M.H., Alpar, M.A. 2004, ApJ, 617, 461

\bibitem[\protect\citeauthoryear{Finger et al.}{1996}]{fin96} 
Finger, M.H., Wilson, R.B., Harmon, B.A. 1996, ApJ, 459, 288 

\bibitem[\protect\citeauthoryear{Fritz et al.}{2006}]{fritz} 
Fritz, S., Kreykenbohm, I., Wilms, J., 
Staubert, R., Bayazit, F., Pottschmidt, K., Rodriguez, J., Santangelo, A. 2006, A\&A, 458, 885

\bibitem[\protect\citeauthoryear{Ghosh et al.}{1977}]{ghoshlamb0} 
Ghosh, P., Pethick, C. J., Lamb, F. K. 1977, ApJ, 217, 578

\bibitem[\protect\citeauthoryear{Ghosh \& Lamb}{1978}]{ghoshlamb} 
Ghosh, P., Lamb, F.K. 1978, ApJ, 223, L83

\bibitem[\protect\citeauthoryear{Ghosh \& Lamb}{1979a}]{ghoshlamb1}
Ghosh, P., Lamb, F. K. 1979a, ApJ, 232, 259

\bibitem[\protect\citeauthoryear{Ghosh \& Lamb}{1979b}]{ghoshlamb2}
Ghosh, P., Lamb, F. K. 1979b, ApJ, 234, 296

\bibitem[\protect\citeauthoryear{Ghosh \& Lamb}{1992}]{ghoshlamb92}
Ghosh, P.; Lamb, F. K. 1992, in X-ray binaries and recycled pulsars, 
Ed. E. van den Heuvel and S.A. Rappaport 
(Kluwer Academic Publishers, Boston), p. 487

\bibitem[\protect\citeauthoryear{Ghosh \& Lamb}{1995}]{ghoshlamb95}
Ghosh, P.; Lamb, F. K. 1995, in Compact stars in binaries, 
Ed. J. van Paradijs, E.P.J. van den Heuvel, E. Kuulkers ù
(Kluwer Academic Publishers, Dordrecht), p.57 


\bibitem[\protect\citeauthoryear{Howe et al.}{1983}]{howe} 
Howe S.K., Primini, F.A., Bautz, M.W., et al, 1983, ApJ, 272, 678

\bibitem[\protect\citeauthoryear{in't Zand et al.}{1998}]{zand98} 
in't Zand, J.J.M., Baykal, A., Strohmayer, T.E. 1998, ApJ, 496, 386

\bibitem[\protect\citeauthoryear{Kaur et al.}{2008}]{kaur08} 
Kaur, R., Paul, B., Kumar, B., Sagar, R. 2008, ApJ, 676, 1184

\bibitem[\protect\citeauthoryear{King et al.}{2007}]{king} 
King, A.R., Pringle, J.E., Livio, M. 2007, MNRAS, 376, 1740 

\bibitem[\protect\citeauthoryear{van der Klis}{1995}]{klis95}
van der Klis, M. 1995, in X-Ray Binaries, 252 

\bibitem[\protect\citeauthoryear{van der Klis}{2004}]{klis04}
van der Klis, M. 2004, arXiv:astro-ph/0410551 

\bibitem[\protect\citeauthoryear{Klochkov et al.}{2007}]{klochkov07} 
Klochkov, D., Shakura, N., Postnov, K., Staubert, R., Wilms, J., Ketsaris, N. 
2006, arXiv:astro-ph/0609276 

\bibitem[\protect\citeauthoryear{Krauss et al.}{2007}]{krauss07}
Krauss, M.I., Schulz, N.S., Chakrabarty, D., Juett, A.M., Cottam, J. 2007, ApJ, 660, 605

\bibitem[\protect\citeauthoryear{Lamb \& Pethick}{1974}]{lamb} 
Lamb, F.K., Pethick, C.J. 1974, in Astrophysics and gravitation; 
Proceedings of the Sixteenth Solvay Conference on Physics, Brussels, Belgium, 
Editions de l'Universite de Bruxelles 1974, p. 135-141. 

\bibitem[\protect\citeauthoryear{Lamb et al.}{1985}]{lamb85} 
Lamb, F.K., Shibazaki, N., Alpar, M.A., Shaham, J. 1985, Nature, 317, 681 

\bibitem[\protect\citeauthoryear{Li \& Wang}{1996}]{liwang2} 
Li, X.-D., Wang, Z.-R. 1996, A\&A, 307, L5

\bibitem[\protect\citeauthoryear{Li \& Wang}{1999}]{liwang} 
Li, X.-D., Wang, Z.-R. 1999, ApJ, 513, 845

\bibitem[\protect\citeauthoryear{Li et al.}{1996}]{liwickra2} 
Li, J., Wickramasinghe, D.T., Rudinger, G. 1996, MNRAS, 469, 765

\bibitem[\protect\citeauthoryear{Li \& Wickramasinghe}{1997}]{liwickra} 
Li, J., Wickramasinghe, D.T. 1997, MNRAS, 286, L25

\bibitem[\protect\citeauthoryear{Lovelace et al.}{1995}]{lovelace} 
Lovelace, R. V. E., Romanova, M. M., Bisnovatyi-Kogan, G. S. 1995, MNRAS, 244, 254 

\bibitem[\protect\citeauthoryear{Makishima et al.}{1992}]{makishima92} 
Makishima, K., Mihara, T., Nagase, F., Murakami, T. 1992, in Proc. 28th 
Yamada Conference: Frontiers of X-ray
Astronomy, ed. Y. Tanaka \& K. Koyama, Frontiers Science
Series (Tokyo: Universal Academy Press), 23

\bibitem[\protect\citeauthoryear{Moon \& Eikenberry}{2001a}]{moon} 
Moon, D., Eikenberry, S.S. 2001, ApJ, 549, L225

\bibitem[\protect\citeauthoryear{Moon \& Eikenberry}{2001b}]{moon01} 
Moon, D., Eikenberry, S.S. 2001, ApJ, 552, L135

\bibitem[\protect\citeauthoryear{Negueruela \& Okazaki}{2001}]{negueruela01} 
Negueruela, I. \& Okazaki, A.T. 2001, A\&A, 369, 108  

\bibitem[\protect\citeauthoryear{Parmar et al.}{1989}]{parmar} 
Parmar, A.N., White, N.E., Stella, L. 1989, ApJ, 338, 373

\bibitem[\protect\citeauthoryear{Parmar et al.}{1999}]{par99} 
Parmar, A.N., Oosterbroek, T., Dal Fiume, D., et al. 1999, A\&A, 350, L5

\bibitem[\protect\citeauthoryear{Petterson}{1975}]{petterson75}
Petterson, J.A. 1975, ApJ, 201, L61

\bibitem[\protect\citeauthoryear{Raichur \& Paul}{2008}] {raichur} 
Raichur, H. \& Paul, B. 2008, preprint (astro-ph/0806.0949) 

\bibitem[\protect\citeauthoryear{Romanova et al.}{2002}] {romanova} 
Romanova, M.M., Ustyugova, G.V., Koldoba, A.V., Lovelace, R.V.E. 2002, ApJ, 578, 420

\bibitem[\protect\citeauthoryear{Romanova et al.}{2003}] {romanova1} 
Romanova, M.M., Ustyugova, G.V., Koldoba, A.V., Wick, J.V., Lovelace, R.V.E. 2003, ApJ, 595, 1009

\bibitem[\protect\citeauthoryear{Romanova et al.}{2004}] {romanova3} 
Romanova M.M., Ustyugova G.V., Koldoba A.V., Lovelace R.V.E. 2004, ApJ, 616, L151
 
\bibitem[\protect\citeauthoryear{Scharlemann}{1978}]{sharl} 
Scharlemann, E.T. 1978, ApJ, 219, 617

\bibitem[\protect\citeauthoryear{Shakura \& Sunyaev}{1973}]{ss73} 
Shakura, N.I., Sunyaev, R.A. 1973, A\&A, 24, 337

\bibitem[\protect\citeauthoryear{Shinoda et al.}{1990}]{shi} 
Shinoda, K., Kii, T., Mitsuda, K., et al 1990, PASj, 42, L27

\bibitem[\protect\citeauthoryear{Soong \& Swank}{1989}]{soong89} 
Soong, Y. \& Swank, J.H. 1989, in X Ray Binaries, the 23rd ESLAB Symposium 
on Two Topics in X Ray Astronomy, p. 617

\bibitem[\protect\citeauthoryear{Steele et al.}{1998}]{steele98}
Steele, I.A., Negueruela,I., Coe, M.J., Roche, P. 1998, MNRAS, 297, L5 

\bibitem[\protect\citeauthoryear{Takeshima et al.}{1991}]{take} 
Takeshima T., Dotani, T., Mitsuda, K., Naga, F. 1991, PASJ, 43, L43

\bibitem[\protect\citeauthoryear{Tamura et al.}{1992}]{tam} 
Tamura, K., Hiroshi, T., Kitamoto, S., et al. 1992, ApJ, 389, 676

\bibitem[\protect\citeauthoryear{Ustyugova et al.}{2006}]{usty06}
Ustyugova, G. V.; Koldoba, A. V.; Romanova, M. M.; Lovelace, R. V. E.
 
\bibitem[\protect\citeauthoryear{Vietri}{2008}]{vietri} 
Vietri, M. 2008, Foundations of high-energy astrophysics, Chicago University press, pp. 325 

\bibitem[\protect\citeauthoryear{Wang}{1987}]{wang87} 
Wang, Y.-M. 1987, A\&A, 183, 257

\bibitem[\protect\citeauthoryear{Wang}{1995}]{wang95} 
Wang, Y.-M. 1995, ApJ, 449, L153

\bibitem[\protect\citeauthoryear{Wang}{1996}]{wang96} 
Wang, Y.-M. 1996, ApJ, 465, L111

\bibitem[\protect\citeauthoryear{Wang}{1997}]{wang97} 
Wang, Y.-M. 1997, ApJ, 475, L135

\bibitem[\protect\citeauthoryear{Wang}{1997b}]{wang97b} 
Wang, Y.-M. 1997, ApJ, 487, L85

\bibitem[\protect\citeauthoryear{Wilson et al.}{1997}]{wilson97}
Wilson, R.B., Scott, D.M., Finger, M.H. 1997, AIP Conf. Proc., 410, 739

\bibitem[\protect\citeauthoryear{Wilson et al.}{2002}]{wilson02}
Wilson, C.A., Finger, M.H., Coe, M.J., Laycock, S., Fabregat, J. 2002, ApJ, 570, 287 
	
\bibitem[\protect\citeauthoryear{Woo et al.}{1996}]{woo} 
Woo, J.W., Clark, G.W., Levine A.M., et al. 1996, ApJ, 467, 811
	
\bibitem[\protect\citeauthoryear{Zane et al.}{2004}]{zane04}
Zane, S., Ramsay, G., Jimenez-Garate, M.A., Willem den Herder, J., Hailey, C.J. 
2004, MNRAS, 350, 506 

\end{thebibliography}
\end{document}